\documentclass[12pt]{article}

\usepackage{graphicx}% Include figure files
\usepackage{amssymb}

\usepackage{times}

\newenvironment{sciabstract}{%
\begin{quote} \bf}
{\end{quote}}

\newcounter{lastnote}
\newenvironment{scilastnote}{%
\setcounter{lastnote}{\value{enumiv}}%
\addtocounter{lastnote}{+1}%
\begin{list}%
{\arabic{lastnote}.}
{\setlength{\leftmargin}{.22in}}
{\setlength{\labelsep}{.5em}}}
{\end{list}}

\title{Signatures of electron fractionalization in ultraquantum bismuth}

\author
{ Kamran Behnia$^{1\ast}$, Luis Balicas$^{2}$ and Yakov Kopelevich$^{3}$\\
\\
\normalsize{$^{1}$ Laboratoire Photons et Mati\`ere (CNRS-UPR5),
ESPCI}\\
\normalsize{10, Rue Vauquelin, 75231 Paris, France}\\
\normalsize{$^{2}$ National High Magnetic Field Laboratory} \\
\normalsize{Florida State University, Tallahassee, Florida 32306, USA}\\
\normalsize{$^{3}$ Instituto de Fisica ``Gleb Wataghin'', UNICAMP}\\
\normalsize{13083-970 Campinas, S\~{a}o Paulo, Brazil}\\
\\
\normalsize{$^\ast$ To whom correspondence should be addressed;
E-mail:  kamran.Behnia@espci.fr.} }

\date{}

\begin{document}

% Double-space the manuscript.

\baselineskip24pt

% Make the title.

\maketitle

% Place your abstract within the special {sciabstract} environment.

\begin{sciabstract}
In elemental bismuth (contrary to most metals), due to the long
Fermi wavelength of itinerant electrons, the quantum limit can be
attained with a moderate magnetic field. Beyond this limit,
electrons travel in quantized orbits whose circumference (shrinking
with increasing magnetic field) becomes shorter than their Fermi
wavelength.  We present a study of transport coefficients of a
single crystal of bismuth up to 33 T, i.e. deep in this ultraquantum
limit. The Nernst coefficient presents three unexpected maxima which
are concomitant with quasi-plateaus in the Hall coefficient. The
results suggest that this bulk element may host an exotic quantum
fluid reminiscent of the one associated with the fractional quantum
Hall effect and raise the issue of electron fractionalization in a
three dimensional metal.
\end{sciabstract}

% In setting up this template for *Science* papers, we've used both
% the \section* command and the \paragraph* command for topical
% divisions.  Which you use will of course depend on the type of paper
% you're writing.  Review Articles tend to have displayed headings, for
% which \section* is more appropriate; Research Articles, when they have
% formal topical divisions at all, tend to signal them with bold text
% that runs into the paragraph, for which \paragraph* is the right
% choice.  Either way, use the asterisk (*) modifier, as shown, to
% suppress numbering.

Electronic properties of bismuth have been extensively studied
during the 20th century. As early as 1928, it was discovered that
its resistivity increases by several orders of magnitude in presence
of a large magnetic field and shows no sign of
saturation\cite{kapitza}. Two years afterwards, studies on bismuth
led to the discovery of quantum oscillations in both
magnetization\cite{dehass} and in resistivity\cite{shubnikov}.
Bismuth was the first metal whose Fermi surface was experimentally
identified\cite{shoenberg1}. Commenting on the exceptional role
played by bismuth in the history of metal
physics\cite{edelman,issi}, Falkovskii wrote in 1968: ``It is
easiest to observe in bismuth the phenomena that are inherent in all
metals.''\cite{falkovskii}

An extremely small Fermi surface and a very long mean-free-path are
what distinguish bismuth from other metals. The Fermi surface
occupies 10$^{-5}$ of the Brillouin zone\cite{edelman}, an order of
magnitude lower than graphite, the closest rival and another
celebrated semi-metal. The mean-free-path at room temperature
exceeds $2\mu m$\cite{pippard}, almost two orders of magnitude
longer than in copper. Due to the low carrier density, the quantum
limit in bismuth can be reached by the application of a magnetic
field as small as 9 T along the trigonal axis. In this limit,
electrons are all pushed to the lowest Landau level and the magnetic
length (the radius of the lowest-energy quantized isolated electron
orbit in a magnetic field) becomes shorter than the Fermi
wavelength. As recently noted\cite{abrikosov}, the quasi-linear
magneto-resistance of bismuth\cite{kapitza} in this limit does not
fit in to the quasi-classical theory of electronic transport. The
last experimental investigation of high-field magnetoresistance in
bismuth, in the 1980s, found no evidence of saturation up to
45T\cite{hiruma}. Interestingly, this was contemporaneous with the
discovery of fractional quantum Hall effect(FQHE)\cite{tsui}. Soon,
the many-particle quantum theory succeeded in providing an elegant
solution to this unexpected experimental finding  \cite{laughlin}.
Today, the FQHE ground state is an established case of a quantum
fluid whose elementary excitations are fractionally charged. Such a
fluid emerges in high-mobility two-dimensional electron systems
formed in semiconductor heterostructures in presence of a magnetic
field exceeding the quantum limit. In contrast with the integer
quantum Hall effect (IQHE), which can be explained in a one-particle
picture, the occurrence of FQHE implies strong interaction among
electrons and their condensation to a many-body quantum
state\cite{yoshioka}.

Very recently, we reported on the giant quantum oscillations of the
Nernst coefficient in bismuth in the vicinity of the quantum
limit\cite{behnia}. The Nernst signal (the transverse voltage
produced by a longitudinal thermal gradient) was found to peak
drastically whenever the Landau level meets the Fermi level.
Otherwise, it is severely damped. This observation was in
qualitative agreement with a theoretical prediction \cite{nakamura}
invoking a ``quantum Nernst effect'' associated with the IQHE. In
this report, we present new measurements resolving distinct peaks in
the Nernst signal deep in the ultra-quantum limit. Measurements of
the Hall coefficient in the same field range reveal a series of
quasi-plateaus extending over a window marked by fields at which the
Nernst peaks occur. These findings raise the issue of the relevance
of the FQHE physics in a clean three-dimensional compensated
semi-metal. They suggest that electron correlations in bismuth are
stronger than what has been commonly assumed and this elemental
metal may host an exotic quantum fluid.

The lower panel of Fig. 1 contains our primary experimental
observation: The detection of three peaks occurring  at 13.3 T, 22.3
T and 30.8 T in the Nernst response [See the supporting online
material for details on measurement technique and sample
characterization]. These three new peaks follow the rich structure
found in the field dependence of the Nernst signal in the previous
study limited to 12 T\cite{behnia} and emerge well beyond the
quantum limit.

The Quantum limit in bismuth is set by the well-known topology of
the Fermi surface in this compensated
semi-metal\cite{edelman,shoenberg2}: An ellipsoid associated with
hole-like carriers around the T-point of the Brillouin zone and
elongated along the trigonal axis and three cigar-like slightly
tilted electron ellipsoids along the L points. The cross section of
the hole ellipsoid perpendicular to the trigonal axis is
$A^{h}_{K}$=0.0608 nm$^{-2}$. For each of the 3 electron ellipsoids,
the corresponding area is $A^{e}_{K}$=0.0836
nm$^{-2}$\cite{shoenberg2} (See Fig.1A). These numbers set the
quantum limit. It is attained by a magnetic field equal to
$B_{QL}=\frac{A_{K}}{2\pi}\frac{\hbar}{e}$ ($\hbar$ is the reduced
Planck constant and $e$ is the charge of electron), that is (6.4 T)
8.6 T for (holes) electrons. When B=B$_{QL}$, the condition
$\lambda_{F}^{\bot}=2\pi\ell_{B}$  is realized : the circumference
of the quantized electronic orbit becomes equal to the Fermi
wavelength ($\lambda_{F}^{\bot}$ is the Fermi wavelength of the
electrons traveling perpendicular to the field and
$\ell_{B}=\sqrt{\frac{\hbar}{eB}}$ is the magnetic length) . In a
two-dimensional system, this corresponds to a Landau level filling
factor of unity. In a three-dimensional system, there is an infinite
degeneracy along the z-axis. Nevertheless, in analogy with the 2D
case, and in absence of an established terminology, we shall use the
expression ``filling factor'' for the ratio
$\nu=(\frac{{2\pi\ell_{B}}}{\lambda_{F}^{\bot}})^{2}$ .

The  Fermi surface cross sections correspond to the low-field
response of the system, however. Strong magnetic field is known to
modify the two Fermi surfaces in order to maintain charge
neutrality\cite{smith,bompadre}. This feature together with Zeeman
splitting leads to a slight enhancement of the quantum limit. It
occurs at 8.9 T, and is marked by the most dramatic peak in
S$_{xy}$. All previous studies\cite{hiruma,smith,bompadre,yang}
converge in detecting a dip in resistivity at 8.85$\pm$0.25 T and
identifying it as the one corresponding to the first Landau level
[See supporting online text for a detailed discussion].

Bismuth is host to surface states quite distinct from the bulk
semi-metal and with a much higher carrier density\cite{hofmann}. It
is very unlikely that their existence is relevant to our
observations. The metallic state resolved by photoemission on the
111 surface [normal to the trigonal] has a Fermi surface whose
radius is of the order of 0.5-3 $nm^{-1}$\cite{hengsberger} and the
expected quantum oscillations would have a frequency range of
100-1000 T.

Therefore, we are brought to conclude that the three new peaks
emerge in the ultraquantum limit. Given the distance between the
four distinct orbits in the reciprocal space, magnetic breakdown is
an unlikely explanation. Moreover, the peaks resolved here do not
display a B$^{-1}$ periodicity. Fig. 2 presents the high-field data
as a function of B$^{-1}$. The peaks are situated at rational
fractions (2/3, 2/5 and 2/7) of the first integer peak. The
low-field data is presented in Fig. 2B. As seen in the panel c of
the same figure, their B$^{-1}$ positions are close to those of the
dips resolved in resistivity (at T=25 mK and for B $<$ 18
T)\cite{bompadre}. Interestingly, in addition to these peaks
(already identified in the previous report\cite{behnia}), there are
two unidentified peaks between $\nu=1$ and $\nu=2$ anomalies and one
between the $\nu=2$ and $\nu=3$. Assuming that $\lambda_{F}$ is
constant between two successive integer peaks, these peaks occur
close to $\nu=4/3$, $\nu=5/3$ and $\nu=5/2$. Fig. 2c summarizes the
position of all Nernst peaks (both integer and fractional) and
resistivity dips. The upward curvature was previously reported and
attributed to the field-induced modification of the carrier
density\cite{bompadre}. This feature would imply a field-induced
change in $\lambda_{F}$. Therefore, the values of $\nu$ for
fractional peaks, which are extracted by linearly extrapolating the
position of adjacent integer peaks are subject to caution. However,
as there is no visible phase transition and the change in
$\lambda_{F}$ is continuous, the extracted values of $\nu$ are not
expected to differ much from the effective ones.

We also performed low-resolution measurements of resistivity and the
Hall coefficient on our sample at 0.44 K. In agreement with the
previous high-field study\cite{hiruma}, resistivity does not show
any strong feature beyond the quantum limit (see supporting online
figure). The Hall response, $\rho_{xy}$ (Fig.3) is strongly
non-linear for fields exceeding 3 T. Above this field, a rich
structure including a sharp peak at 9.8 T is resolved. At still
higher fields, i.e. in the ultra-quantum limit, a succession of fast
and slow regimes in the field-dependence of $\rho_{xy}$ is visible.
Comparing the relative position of these quasi-plateaus and the
Nernst peaks, one clearly sees that each Nernst peak occurs between
two successive Hall plateaus in agreement with the theoretical
prediction invoking IQHE\cite{nakamura}.

Below 3 T, the slope of $\rho_{xy}$ yields R$_{H} =1.5\times 10^{-5}
m^{3}/C$ corresponding to a carrier density of $3.9\times 10^{17}
cm^{-3}$, only 30 percent larger than the hole carrier density. In
the high-field regime, $\rho_{xy}$ is $0.129 \Omega cm$ at $\nu
=1/2$ and $0.158 \Omega cm$ at $\nu=1/3$. Therefore, when the
filling factor passes from 1/2 to 1/3, $\rho_{xy}$ jumps by
$0.029\Omega cm$ . Let us recall that, in a 2D electron gas in the
FQHE, the magnitude of $\rho^{2D}_{xy}$ at filling factor $\nu$,
would be $ h/(\nu e^{2})$ and the passage from the $\nu=1/2$ to
$\nu=1/3$ plateau would lead to a jump of $h/e^{2}=25.8k\Omega$ in
$\rho^{2D}_{xy}$. Since in our 3D system, the jump in  $\rho_{xy}$
is 11.2 nm times $h/e^{2}$, it is tempting to consider the bulk
crystal as an assembly of coherent 2D sheets each 11.2 nm-thick.
This length scale is very close to  $\lambda_{F} ^{\|}$ =14 nm, the
Fermi wavelength of holes along [the trigonal axis and] the magnetic
field and much longer than the atomic distance between layers
($\sim$ 0.2 nm).

These observations  raise the issue of relevance of FQHE physics to
bulk bismuth. The former is found in the context of high-mobility,
two-dimensional and interacting electronic systems. To what extent,
electrons in bismuth qualify for these attributes? The electronic
mobility is undoubtedly large enough. In our crystal, it exceeds by
two orders of magnitude, the mobility of the GaAs/AlGaAs sample in
which the FQHE was discovered in 1982\cite{tsui}(See supporting
online text). It is true that bismuth is not commonly considered as
a strongly interacting electron system. However, the very low level
of carrier density undermines screening and favors Coulomb
repulsion. Since electrons in bismuth and in GaAs are comparable in
their concentration and in their effective mass, it is reasonable to
assume that Coulomb interaction is sufficiently strong to allow
electron fractionalization.

The most serious obstacle for the realization of FQHE physics in
bulk bismuth is dimensionality. We note that IQHE (also a purely 2D
effect) has already been observed in an anisotropic 3D electron
gas\cite{stormer3} as well as a number of bulk systems (See
supporting online text) and  3D FQHE  has been a topic of
theoretical investigation\cite{balents}. Bismuth, with its weak
anisotropy, was recently proposed as a candidate to exhibit the
quantum spin Hall effect\cite{murakami}. However, according to our
preliminary studies, in the vicinity of the quantum limit, the
in-plane conductivity in bismuth is orders of magnitude lower than
the perpendicular conductivity along the field axis. Instead of
being an assembly of weakly-coupled 2D sheets perpendicular to the
field, the system is closer to to a set of 1D wires oriented along
the field\cite{biagini}. To the best of our knowledge, there is no
appropriate theoretical frame for in-plane transport in such a
context. Thus, bismuth in the ultraquantum limit emerges as an
experimental playground for two distinct routes towards the electron
fractionalization, the FQHE and the field-induced Luttinger
liquid\cite{biagini}.

Several other questions remain. What happens to the electron-like
carriers? Intriguingly, for this field orientation, the three
electron ellipsoids have been invisible in all studies of quantum
oscillations.  The complex structure of $\rho_{xy}(B)$ for B $< $14
T suggests competing responses of electrons and holes. However, in
the ultra-quantum limit, there is no definite signature of their
presence in spite of the larger total area occupied by the three
electron ellipsoids. This may be a consequence of their lower
mobility. The absence of strong features in the raw $\rho_{xx}(B)$
data is also intriguing. In 2D systems, the quantum Hall plateaus
are associated with absence of dissipation and vanishing
longitudinal conductivity. In bulk systems showing IQHE, plateaus in
$\rho_{xy}$ are concomitant with minima in $\rho_{xx}$.  Extensive
high-resolution studies of resistivity at lower temperatures on
cleaner samples may help to identify the source of resistive
dissipation at high fields. The 80 years-old mystery of
magneto-resistance in bismuth\cite{kapitza,abrikosov} needs fresh
experimental and theoretical attention.

\begin{scilastnote}
\item  We thank G. Boebinger, P. M. Chaikin, T. Giamarchi and A. MacDonald for useful discussions and
E. Palm and T. Murphy for their assistance. KB  is supported by
DGA/D4S and ANR (ICENET project) and acknowledges a NHFML-VSP
fellowship. LB is supported by the NHMFL in-house research program.
YK is supported by CNPq and FAPESP. Research at the NHMFL is
supported by NSF Cooperative Agreement No. DMR-0084173, by the State
of Florida, and by the DOE.
\end{scilastnote}
\clearpage

\noindent {\bf Fig. 1.} A) Cross section of the hole-like (center)
and electron-like ellipsoid Fermi surfaces for a field applied along
the trigonal axes.  B) The field dependence of the Nernst signal at
different temperatures. The three
peaks resolved in the ultraquantum limit are indicated by arrows.
\\

\noindent {\bf Fig. 2} A) The high-field Nernst plotted as a
function of B$^{-1}$. Dotted lines represent
$B^{-1}_{p/q}=\frac{p}{q}\times0.1118 T^{-1}$. Note the position of
the three peaks respective to these lines. The inset shows the
thermal broadening of the peak at highest field. B) Quantum
oscillations observed in the low-field data\cite{behnia}. Identified
peaks correspond to integer filling factors of the hole pocket. As
seen in the inset, the B$^{-1}$ positions of two previously
unidentified peaks are very close to 4/3 and 5/3 filling factor. C)
The B$^{-1}$ position of the Nernst\cite{behnia} and
resistivity\cite{bompadre}) anomalies $vs.$ the filling factor
associated with them. For determination of fraction filling factors
see text.
\\

\noindent {\bf Fig. 3.} The field-dependence of the Hall
resistivity. A complicated behavior in the intermediate field range
precedes quasi-plateaus in the ultra-quantum limit. The Nernst
response at 0.83 K is also shown. The quantum limit is marked by a
thick red line. Two dotted lines mark $\nu$ (see text). All other lines
are guides to eye.

\begin{figure}
{\includegraphics[width=14cm]{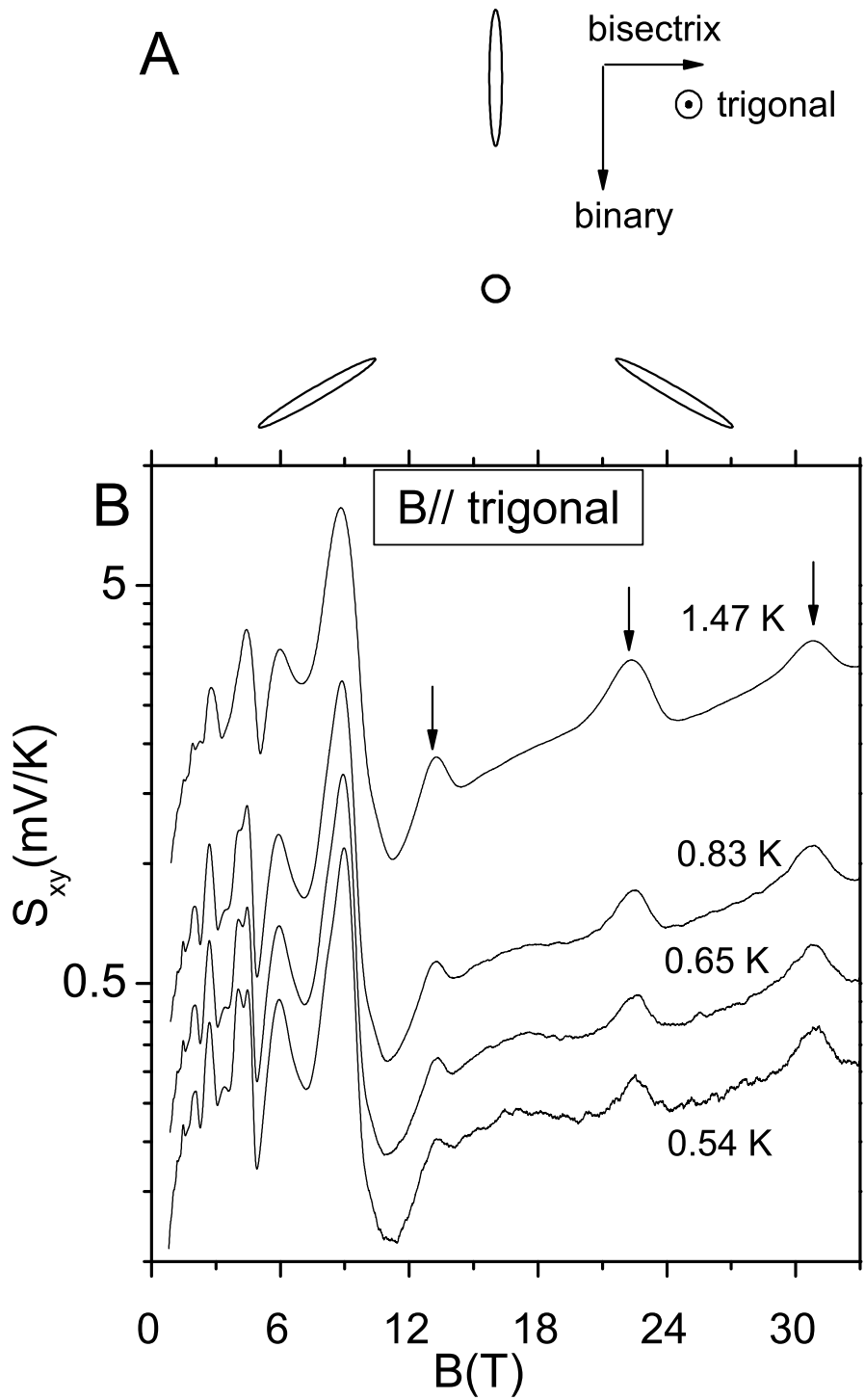}}\caption{Behnia \emph{et
al.}}
\end{figure}
\begin{figure}
{\includegraphics[width=14cm]{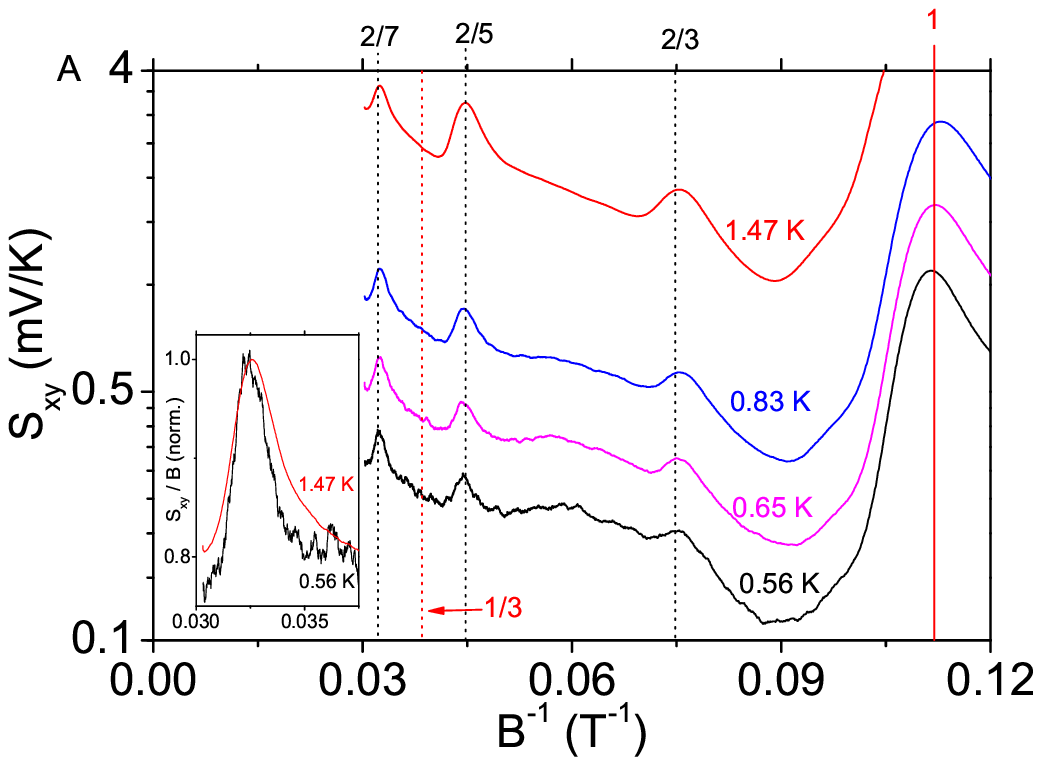}}
{\includegraphics[width=14cm]{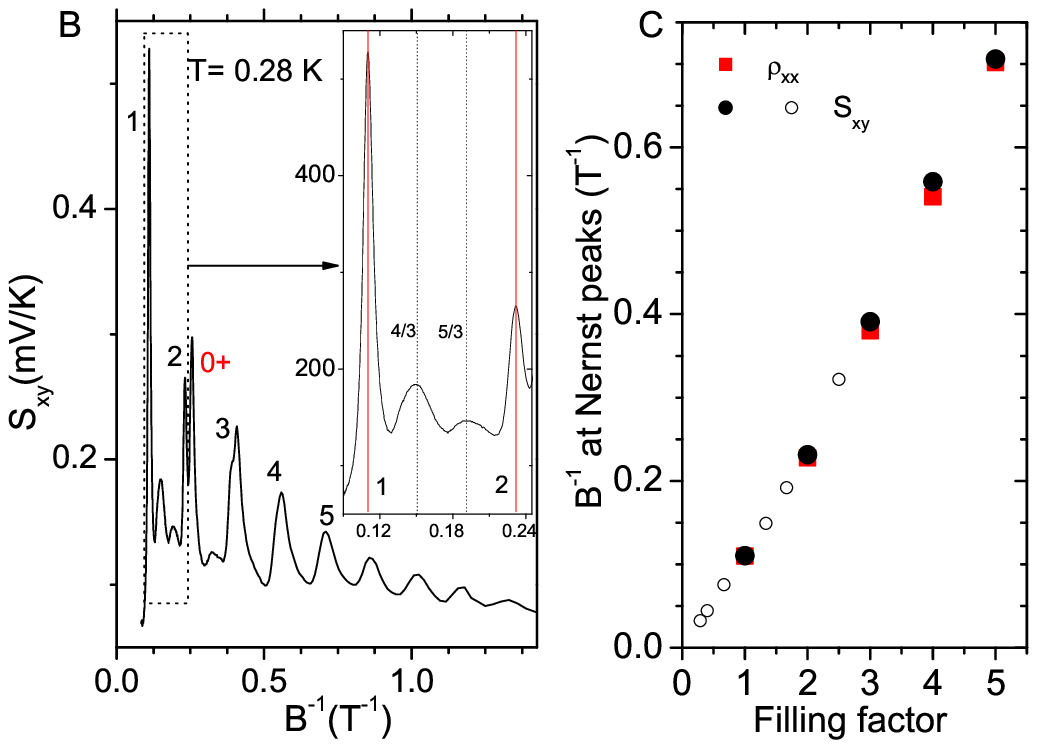}} \caption{Behnia \emph{et
al.}}
\end{figure}
\begin{figure}
{\includegraphics[width=14cm]{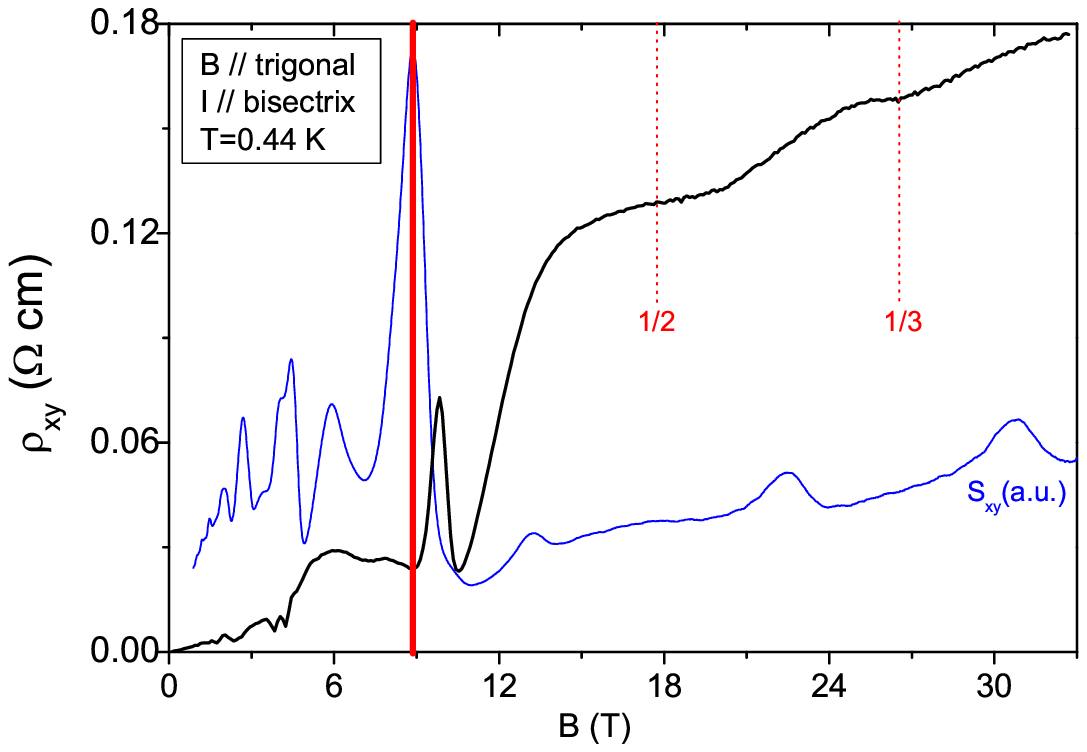}}\caption{Behnia \emph{et
al.}}
\end{figure}

\end{document}